# The dynamics of costly information sharing, falsification and accuracy

**Nathan Berg\*, Chunyu Chen\*\* and Murat Kantarcioglu\*\*\***[1]

## Introduction

Those who study organizational failure commonly diagnose the root cause as breakdowns in information sharing. A vivid example is the failure of US intelligence agencies to share information that, in retrospect, might have prevented 9/11. This interpretation holds that the FBI and CIA could have shared important leads to prevent the attacks but instead responded to what those agencies perceived to be *tournament incentives* under which only one agency within the broader security and threat-prevention apparatus of the US government could expect to be rewarded (financially and otherwise) for stopping an impending threat. In this view, misaligned incentives led to "turf issues" (i.e., parochialism reflecting incompatible objectives between subsidiary and parent organizations) which caused the quantity and quality of information sharing to fall significantly short of the levels necessary for achieving the parent organization's primary objective.

Insufficient information sharing as a root cause of failure to prevent the attacks featured prominently in rationales given by those who then pushed for the creation of the Department of Homeland Security (DHS) in 2002 and passage of the Intelligence Reform and Terrorism Prevention Act (IRTPA) of 2004. The dramatic reorganization of 22 agencies into the DHS and implementation of IRTPA were carried out by officials who frequently stated the goal of transforming incentives to better facilitate cooperative rather than rival motives among subsidiary agencies when making decisions about information to be shared. This goal of transforming incentives led to the creation of new administrative roles, such as Chief Information Sharing Officer for the FBI, the Information Sharing Council to advise the US President, and the Information Sharing Environment Program Manager appointed by the US President.[2] New private organizations appeared as well. For example, the Information Technology Information Sharing and Analysis Center (IT-ISAC), according to its website (https://www.it-isac.org/), works with intelligence agencies on "homeland security issues," seeking legislation to limit liability when private firms share information with government agencies and greater access to classified information for firms in a position to help defend against attacks in cyberspace and beyond. The claim that incentive effects in the context of information sharing are important enough to justify dramatic organizational overhaul (e.g., creation of the Department of Homeland Security and new patterns of wider disbursement of funding for anti-terror programs since 9/11) draws on surprisingly scant empirical research and behavioral testing on information-sharing responses to cooperative versus tournament incentives. This paper attempts to fill that gap.

\* Nathan Berg, Associate Prof. of Economics, email: prof.berg@gmail.com
\*\* Chun-Yu Chen, Ph.D., email: ccy216@gmail.com
\*\*\* Murat Kantarcioglu, Associate Prof. of Computer Science, email: muratk@utdallas.edu
This work was partially supported by Air Force Office of Scientific Research MURI Grant FA9550-08-1-0265.

---

[2] The diagnosis of information-sharing failures as a root cause of 9/11 appears, for example, in the executive summary of the FBI's 2011 Information Sharing Report: "Since September 11, 2001, the FBI has shifted from a traditional crime-fighting agency into an intelligence-led, threat-driven organization, guided by clear operational strategies. Today's FBI is focused on predicting and preventing the threats we face while at the same time engaging with the communities we serve. This shift has led to a greater reliance on technology, collaboration, and information sharing" (FBI, 2011, accessed January 18, 2013, http://www.fbi.gov/stats-services/publications/national-information-sharing-strategy-1/national-information-sharing-strategy). The Centers for Disease Control and Prevention (CDC) similarly issues reports on and allocates substantial administrative resources regarding its information sharing capability (CDC, 2013).

Wageman and Baker (1997) demonstrated incentive effects on group cooperation using task designs with varying degrees of interdependence. Undertaking to formalize and extend the seminal paper of Wageman and Baker to the context of information-sharing behavior, a companion paper of ours proposed a one-shot information-sharing game in which zero sharing is the inefficient Nash Equilibrium under tournament incentives and efficient (i.e., Pareto-Optimal) sharing is the Nash Equilibrium under cooperative incentives. The present paper uses that information-sharing game to elicit new empirical information about experimental participants' responses to shifts in incentives in repeated interaction. The theoretical model's unequivocal predictions apply to a one-shot interaction. In contrast, the conditions under which stable patterns of information sharing emerge in repeated interaction, which is more characteristic of the organizational settings the model is intended to represent, are more subtle and in general non-unique. This motivates the empirical study of repeated information-sharing behavior.

The results provide confirmatory support for the frequent finding of counterproductive effects of competition on task performance reported by Campbell and Furrer (1995) and numerous others, extending that finding to the context of information-sharing tasks. In addition to intelligence agencies such as the FBI and CIA, other interpretations[3] of the information-

---

[3] Links between information sharing and organizational success have numerous antecedents that include a broad range of applications outside the context national security and intelligence sharing. Dawes (1998) describes barriers and risks to information sharing within government organizations (that have no explicit ties to intelligence gathering ) and suggests ideas for enhancing transmission of information. Connelly, Zweig, Webster and Trougakos (2012) detail challenges that organizations face in facilitating knowledge transfer; they introduce a

sharing game include: multinational firms transmitting cost and demand data among subsidiaries; marketing and strategy tasks requiring aggregation of consumer data from multiple locations; scientific networks attempting to synergistically aggregate information from different labs or research groups; transfer pricing as a signaling mechanism for coordinating outputs of intermediate and final goods among geographically dispersed divisions; and supply chain management.

Given the many excellent studies of the effects of incentive schemes on cooperation and organizational performance that the organizational psychology literature already provides, it is worth drawing explicit links to closely related predecessors that motivate this paper's approach while emphasizing what is new. Previous papers usefully characterized information-sharing decisions as public goods games (Whitley, 2003; Cabrera and Cabrera's, 2002; Tsai, 2002; and Monge et al., 1998). This insight guided our attempts to formalize sub-optimal informational sharing in a model with payoff parameters that could capture contrasting cooperative and tournament incentives, and then empirically measure: how much information people share under

---

quantitative measure of *knowledge hiding* that captures more richly what this paper attempts to capture with the binary event of "falsification"(i.e., sharing information that one knows to be false to another member of the organization). The supply chain management literature presents numerous models in which information sharing is the main object of study (Lee, So and Tang, 2000; Chen, 2003; Fiala, 2005; Li and Lin, 2006; Zhou and Benton, 2007; Sahin and Robinson, 2007). Additional studies of information sharing to which the information-sharing game can be applied includes a considerable breadth of important papers in transfer pricing, bilateral negotiation, and marketing and strategy involving decision makers in multiple locations (Ackelsberg and Yukl, 1979; Dejong, Forsythe, Kim, and Uecker, 1989; Chalos and Haka, 1990; Ravenscroft, Haka, and Chalos, 1993; Luft and Libby, 1997; Avila and Ronen, 1999; Ausubel, Cramton, and Deneckere, 2002; Fatima, Wooldridge, and Jennings, 2005; Sawers and Liao, 2005; Cheng and Hsieh, 2009).

different incentives, their propensities to falsify information, and rates of success in achieving accurate aggregations of data their organizations required them to know.[4]

Recent thinking on organizational incentives have, in some important settings, moved away from binary taxonomies for the incentives that influence cooperation (e.g., cooperative versus individualistic) and toward a mixed approach acknowledging that individual and collective rewards can function complementarily (Wagner, Humphrey, Meyer and Hollenbeck, 2012). Our experiment exposes all participants to both cooperative and tournament incentives in randomized order, and our data analysis looks for evidence of combined exposures improving performance or otherwise giving rise to distinct behavioral responses.

---

[4] Whitley (2003) emphasizes the distinct dimensions of autonomy and authority that generate a wide variety of institutional forms which organizations use to solve collective action and other coordination problems. Cabrera and Cabrera's (2002) study of "knowledge-sharing dilemmas" describes coordination and incentive problems surrounding information sharing as a public goods game amenable to shifts in incentive schemes to motivate improved quantities and quality of information sharing. Their proposed framework is perhaps closest in spirit to the present paper's. The formal model that formed the basis for this paper's experimental design follows Cabrera and Cabrera's (2002) description. Importantly, Cabrera and Cabrera (2002) propose that internal transmission of information is costly (in the eyes of individuals and subsidiaries within an organization). By introducing a monetary cost of information sharing, we undertake once again to capture Cabrera and Cabrera's notion of costly information transmission that provides the central economic trade-off between expected benefits versus costs of information sharing. Monge, Fulk, Kalman, Flanagin, Parnassa and Rumsey (1998) propose a theoretical model that provides additional motivation for the formulation of information sharing as a public goods or collective action game. Our experimental design facilitates empirical tests for those incentive effects predicted by theory that are straightforward to implement as hypothesis tests of algebraic restrictions on regression coefficients. Tsai (2002) uses socio-metric measures in a hierarchical modeling framework to study incentive effects on information sharing within a multi-unit organization where subsidiaries compete for market share and internal resources.

Compared to studies based on rich ethnographic and interview data that provide detailed accounts of the processes and institutions that enable information sharing in geographically dispersed teams (Baba, Gluesing, Ratner and Wagner, 2004), the theory and experimental protocol in this paper may strike some readers as psychologically narrow given its reliance on monetary payoffs as the mechanism for modeling incentives. As discussed in subsequent sections of this paper, however, the data reveal behavioral patterns that cannot be explained solely in terms of self-interested payoff maximization.[5] The simplicity of the reward structure (based on payments for accuracy in anonymous interactions sustained over four sets of 16 experimental rounds) generates sharp behavioral findings implying the existence of important motives outside the standard rational actor model. The design's simple payoff structure (not embedded in richly contextualized organizational narrative) also permits a rather broad range of interpretations as mentioned already.

---

[5] The present paper's simple payoff structure should be interpreted as complementary with more complex sociological theory and empirics on multi-national organizations. For example, Vora and Kostova (2007) report empirical findings related to ours based on constructs from social identity theory that influence subsidiary-parent cooperation and knowledge transfer. Interesting sociological modulators of incentive effects are reported in Quigley, Tesluk, Locke and Bartol (2007). Eby and Dobbins (1997) investigate psychometric correlates of self-reported collectivism at individual and group levels. Golden and Raghuram (2010) provide empirical detail about social influences on teleworkers' information sharing. Shared cognition of course involves much richer informational exchange than the mere transmission of data. Nevertheless, similar sets of issues regarding how to measure information sharing and evaluation of its effects arise in Salas and Cannon-Bowers (2001) and Ensley and Pearce (2001). Intra-organization information sharing is a longstanding research focus in the management information science literature (e.g., Barrett and Konsynski, 1982). Haas and Park (2010), and Rotolo and Petruzzelli (2012) investigate the problem of academic researchers withholding information and the role of both peer effects and incentives in explaining observed patterns of information sharing--yet another organizational domain where the model in this paper and its focus on tension between cooperative versus individualistic incentives is applicable.

Another goal of the experimental analysis is to investigate the persistence of previous incentive schemes after they have been replaced by new ones, interpreted as historical (or path-) dependence when new incentive schemes are introduced. Managers would want to understand the potential for path-dependence to affect the magnitudes and timing of behavioral responses to new incentives, as well as the need for other supportive interventions that may be required to successfully transition away from entrenched non-cooperative organizational culture rooted in narrowly individualistic reward schemes from the past. In the other direction, persistence of cooperative information sharing during temporary spells in which rewards become more individualistic could add to an organization's understanding of the value of its cooperative institutions. Nalbantian and Schotter's (1997) work demonstrates persistence of cooperation after incentive schemes shift from team-based to individual rewards. In contrast, Harbring (2010) shows persistence of non-cooperative behavior when transitioning from environments with dysfunctional competitive incentives toward reward schemes designed to facilitate greater cooperation.[6] The experimental design adopted in this paper enables formal tests of order-of-treatment effects that, while limited by the brief period of interaction in the experimental lab, provides suggestive evidence that such phenomena are likely to exist.

---

[6] Economists have also studied information sharing as a function of characteristics of good or demand behaviour (i.e., complements versus substitutes) (Cason, 1994) and the role of information sharing in achieving optimality (Vives, 1984; Li, 1985). In transfer pricing applications that require extensive information transmission among subsidiaries of a multinational organization, inaccurate information received from others leads to the so-called transfer price expectations gap, which impedes progress in bargaining and leads to suboptimal

Another potentially important channel through which the results in this paper translate into insights for designing incentive mechanisms that support information sharing is Gigone and Hastie's (1993) "common knowledge effect." Gigone and Hastie showed that information distributed to all group members influences judgments more than information that is only partially distributed does. This effect has potential to influence the weight that subsidiaries place on different pieces of information in their possession, as a function of how widely distributed that information is. The common knowledge effect might therefore interfere with an organization's information sharing by distorting attention away from privately held information or, alternatively, could be used strategically to induce subsidiaries to place greater weight on particular pieces of (widely distributed) information. In addition to cooperative and tournament incentive treatments, our experimental design also includes treatments varying the size of participants' information endowments. Low-information treatments partition the organization's database into equal, non-overlapping information endowments. High-information treatments provide participants with larger information endowments that overlap. Treatments that vary the quantity of information that each player initially possesses reveal a paradoxical less-is-more effect: when everyone has less information, they share more. The final section of this paper

---

outcomes (Linhart and Radner, 1992; Chang, Cheng, and Trotman, 2008). Chalos and Haka (1990) demonstrate that subsidiaries frequently act on incentives to advance the subsidiary's interests at the expense of others (and the entire organization's profits), by withholding information and exploiting the parts that remain private.

briefly discusses the cautionary implications of this finding with regard to the apparent "more-is-better" approach to collecting information in many real-world organizations including the DHS.

**Hypotheses**

Given the dramatic reorganizations mentioned in the introduction seeking to improve information sharing, it would seem that studying how people making repeated decisions about how much information to share, whether to share honestly, and whether to trust what is shared by others, are clear research priorities. This paper seeks a descriptive account of these information-sharing behaviors based on experimental data and game theory, with which to measure incentive effects that affect organizational performance. The experimental design based on the information-sharing game makes the assumption that organizational performance depends on the objective accuracy that teams are able to achieve after making decisions about how much information to share and whether to deliberately share false information.

The data reporting in the following sections seek to answer five basic questions about incentive effects on the dynamics of information sharing in an otherwise neutral environment without contextualizing narrative as part of a particular organization, without priming language regarding team and organizational membership that might encourage cooperation, and without face-to-face contact or communication other than those information-sharing decisions that are passed to other players electronically[7]:

---

[7] Although common in some subfields of behavioral science such as experimental economics, neutral-language experimental design as a methodological choice attracts heated debate. In psychology, context-neutral design is criticized by those arguing from Brunswikian (and other) methodological viewpoints that rich contextualization improves data quality by reducing noise from unobservable framing in the minds of experimental participants. The goal of our

(1) Do cooperative incentives that reward all team members for success by any one member generate optimal or near-optimal levels of information sharing?

(2) In contrast, do tournament incentives rewarding only a single person who first achieves success (or a single division within a multi-divisional organization) induce people to share no information as theory predicts?

(3) How do cooperative versus tournament incentives affect team members' propensities to falsify information, even though doing so is costly and never an own-payoff-maximizing strategy?

(4) Do incentive schemes have important effects on the objective accuracy of the aggregated databases that individuals achieve?

(5) Does the *ex-ante* quantity of information that individuals (or divisions) are endowed with prior to making decisions on information sharing and falsification affect quantities shared, rates of falsification and accuracy?

## Experimental Design

The experiment is designed to follow a game-theoretic model whose expected payoff matrices and Nash equilibria are presented in Appendices 1 and 2. Two players representing divisions of a multi-location organization make three simultaneous decisions: the number of database entries to share, referred to as passing or transmitting information to another player representing a distinct division within the organization; whether to falsify the information passed to other players; and whether to distrust (as opposed to use at face value) the information

---

experimental design is to provide clean and simple tests of incentive effects on quantities of information shared, rates of falsification, and accuracy.

received from other players. Each player receives an information endowment containing a partial list of entries from the organization's complete database, which is known to be accurate but distributed in discrete information endowments allocated across different players or divisions. Payoff maximization requires each player to gather enough information to form an accurate view of the organization's complete database. This database describes a mapping from characteristics into "targets," which can be thought of as a dataset on which the organization's success requires a regression analysis to uncover the true conditional expectation represented by the complete set of database entries. If any person had access to all players' information endowments, then he or she would have a perfect (or at least the best possible) estimate of the regression on which the organization's success depends.

To keep the information-gathering task easy to understand, each database *entry* consists of only three binary variables, which is simply a row vector associating two observable characteristics (i.e., cues or predictors) with a y-variable coding target/non-target status. For example, Table 1 shows an example of a complete database (which no player has access to prior to sharing information) comprised of four entries (i.e., row vectors referred to as database entries) that record nationality, language spoken, and target/non-target status. The complete database in Table 1 indicates that a non-Pakistani Pashto speaker is a target and that all other combinations of Pakistani and Pashto-speaking characteristics are not targets.

In low-information treatments (A and B), each player receives a non-overlapping information endowment of two database entries drawn from among the four entries comprising the complete database. In high-information treatments (C and D), each player receives four entries drawn randomly with replacement. There is common knowledge that both players' *ex-ante* information sets overlap with very high probability. Players earn financial rewards (with either cooperative bonuses or tournament bonuses as described below) on the basis of a test of the accuracy of their *view* (i.e., guess, conjecture, or estimate) of the complete database. Financial bonuses are only awarded for submitting a perfectly accurate view of the database.

Information sharing incurs an explicit financial cost of $1 per database entry shared with the other player. The monetary cost of sharing information is intended to represent tangible, strategic and/or psychological costs (as described, for example, in Cabrera and Cabrera, 2002).[8] Monetary cost per-information-unit-shared is an experimental design parameter that generates easy-to-measure trade-offs between explicit costs of sharing information (which are known as common knowledge) and the benefit of higher expected payoffs from cooperatively sharing information. Although costly, sharing information is the only way (in the model and experiment) to guarantee that one or more players can achieve an accurate view of the database and reveal, for the benefit of the parent organization, how observable characteristics map into target/non-target status.

---

[8] In contrast to this paper (in which the monetary costs of information sharing are an exogenous theoretical and experimental design parameter), Dyer and Chu (2003) investigate trust as the variable that drives variation in perceived transaction costs of information sharing. Information-sharing costs could be varied in our experimental framework by introducing new treatments with different cost structures for passing information. The predictions of the theoretical model (of the stage game in Appendices A and B) are highly robust to variation in this cost parameter, however. That means the equilibrium predictions are stable over a very wide range around the value of $1 per database entry. Fulk, Heino, Flanagin, Monge and Bar (2004) report evidence indicating the importance of the costs of sharing information in explaining observed decisions about sharing information. Hansen's (2002) work focuses on the fixed costs of maintaining information-sharing networks rather than variable costs (that depend on the number of database entries shared as in the present paper). Fixed costs, within bounds, could be included in our model without changing its qualitative predictions.

*Cooperative and Tournament Bonus Payments*

The experimental design's main treatment variable is cooperative versus tournament bonus payments. In the cooperative treatments (A and C), both players receive bonus payments of $12 if either player achieves an accurate view of the database. In the tournament treatments (B and D), the first player to achieve an accurate view of the database receives a bonus of $24.[9] As described earlier, treatments also vary the number of database entries in players' information endowments. In low-information treatments (A and B), each player's information endowment contains two database entries that do not overlap with the other player's endowment. In high-information treatments (C and D), the two players' endowments consist of four entries that are very likely to contain repeats (since they were drawn with replacement) and overlap with the other player's information endowment.

Treatments are sequenced so that around half of participants are exposed to cooperative Treatment A first (according to the sequence A-B-C-D) and half are exposed to a tournament Treatment B (sequence B-A-D-C). This generates information from which subsequent data analyses extract evidence about path dependence. This evidence will be referred to as order-of-treatment effects, which show whether previous incentive schemes produce measurable spillovers influencing present information-sharing behavior after new incentive schemes have been introduced. Participants are paid at the conclusion of the experimental session for a randomly chosen round determined by a participant rolling two dice that generate random numbers from 1 to 64.

---

[9] Gibson, Waller, Carpenter and Conte (2007) demonstrate the importance of speed and timing within multi-national organizations, providing motivation for the introduction of timing-contingent bonuses in Treatments B and D. Although it would have been possible for speed to have determined which participant received the bonus payoff in any of the 1,600 pair-rounds we observed, it was accuracy rather than speed that determined which player got the bonus in the overwhelming majority of cases (more than 97% of 1,600 group observations, pooling over the two tournament treatments).

*Decision 1: #Shared*

When information endowments are comprised of 2 entries, the range of choices for #Shared is 0, 1 or 2. When information endowments are comprised of 4 entries, the range expands to integers from 0 to 4. The two players move simultaneously, which means they cannot condition their decisions on the contemporaneous (i.e., same-round) decisions of the other player.[10]

*Pairing and Re-Pairing of Players for 16-Round Repeated Interaction in Each Treatment*

Participants are randomly paired for each 16-round treatment. During the 16 rounds of each treatment, the composition of each pair is fixed (i.e., players stick with the same "other player.") Before the next treatment begins, participants are re-paired with a new person in the role of "other player," fixing the participants in each pair for the next 16 rounds of the following treatment. Therefore, each treatment is a repeated interaction with the same "other person." The t-values presented in all subsequent tables are corrected to allow for within-pair dependence over the multiple pair-rounds in which group outcomes are observed. Clustering on pair ids for each treatment allows for within-pair correlation among error terms over the rounds in which each pair interacted. Clustering with robust standard errors produces the larger (more conservative) standard errors and smaller t statistics reported subsequently in the regression results.

---

[10] Appendix 3 shows a screen shot of the z-tree interface eliciting #Shared.

*Decision 2: Falsification*

After choosing the quantity of information to share, players next face a z-tree computer screen requiring them to "fill in," using a mouse, all database entries they wish to share. Although participants can always see the entries in their endowments, they are not presented with a list of those entries and prompted to simply point and click. Instead, the computer interface requires those who have chosen to share to do the electronic equivalent of writing database entries by hand. Each of the three components of every database entry ($x_1$, $x_2$, $y$) must be selected using toggle switches that allow for any possible database entry to be shared rather than restricting to the list of entries one knows to be true (i.e., the entries received in one's information endowment).[11]

With the goal of avoiding normative priming, refraining from tacit suggestions or any expressions of judgment about the right way to share information, the pre-experiment training makes clear to participants that they are not required or expected to pass only database entries they know to be true (i.e., the ones received in their information endowment). Because each player sees his or her information endowment (which remains visible on the screen while filling in the database entries to be shared), it is very easy for participants to have full self-awareness of, and control over, the decision to pass an entry he or she knows to be untrue.

Falsification is coded as a binary outcome = 1 if an individual passes one or more database entries that do not match any of the entries in his or her information endowment. A player can falsify and accidently share an entry that is in fact true. Thus, falsification here is defined solely in terms of what the person doing the sharing knows at the time the information is shared. Falsification is obvious, easy to avoid, and therefore interpreted as deliberate whenever it is occurs. Since information is costly and success depends on sharing, falsification is never a payoff-maximizing strategy.[12]

---

[11] Appendix 4 shows a screen shot of a player who has already decided to share two database entries "filling in" each component of the entries they intend to share.

*Test of Accuracy, Decision 3 (Distrust) and Bonus Payments*

Once the shared database entries (if any) are received on both sides, each player is tested on the accuracy of his or her view (i.e., estimate or guess) of the complete database by describing the target- or y-values associated with each of the four vectors of x-characteristics in the database.[13] While participants undertake this test of accuracy, their information endowments are continuously displayed onscreen together with any database entries received from the other player, which are clearly indicated as such. For example, if player 1 has a two-entry endowment and receives two non-false entries from player 2, then player 1 will achieve accuracy in his or her view of the database (so long as player 2 did not falsify and player 1 does not distrust the true information that was received).

---

[12] Lau and Cobb (2010) distinguish calculus-based versus relationship-based trust. There is tension between these (and the potential presence of both) aspects, because the standard own-payoff-maximizing prescription would, if common knowledge, lead both players to predict zero falsification from other players. This is true in both cooperative and tournament incentive treatments. Once a player sees that the other player falsifies (even occasionally), this already reveals that the other player is not best-responding or playing a Nash Equilibrium strategy. More can be said, however, because there is no belief that can rationalize an own-payoff-maximizing to falsify. It is always cheaper to withhold information than pass false information. This means that when the other player is observed to falsify, then his or her partner can logically conclude that the person who falsified is not a rational own-payoff maximizer or otherwise believes in a behavioral (i.e., non-maximizing) theory of the other player. For example, a player might believe the other player pays attention to the number of entries shared but cannot detect when a false piece of information is shared. Under this behavioral theory of the other player, one could reason that the other player will retaliate harshly if I withhold information, which would imply that it costs less to sneak a false piece of information (thereby reducing the chance that others receive the bonus payment) in tournament treatments. Whether this could be an effective strategy depends crucially on: whether the passing of false information is detected; beliefs about that event's probability; and beliefs about likely reactions from the other player. Butler's (1999) experimental study of the quantity of information shared as a function of trust versus distrust links trust to greater quantities shared and profits.

We can only partially observe distrust. Behavioral (rather than introspective or subjective) distrust occurs when an entry is received from the other player and the receiver then contradicts the target value of the received entry when performing the test for accuracy. For example, if a player receives the database entry consisting of the 1x3 row vector (Pakistani, Pashto, not target) but indicates the row vector (Pakistani, Pashto, target) on his or her accuracy test, which matches on $x_1$ and $x_2$ but mismatches on the y-value, then this response would be classified as an instance of behavioral distrust. A limitation of this measure can be seen when Player 1 distrusts what was received from Player 2 and randomly guesses one or more entries (in the accuracy test) that happen to have been shared by Player 2. In that case, Player 1's guess coincides with what Player 2 shared and, therefore, there is no observable instance of distrust to record. This example shows the incompleteness of the observational definition of distrust, which fails to record any such unobservable instances of distrust.

---

[13] A screenshot of the accuracy test is provided in Appendix 5. The accuracy test can be represented as filling in the following four blanks

$$(0, 0, \_), (0, 1, \_), (1, 0, \_), \text{ and } (1, 1, \_),$$

while the information endowment and any information received from the other player are simultaneously displayed on screen.

One way to be *in*accurate is to have received too few entries from the other player so that guessing is required and then guess incorrectly. Another way to be inaccurate is to distrust a received entry which is in fact true. The third way to be inaccurate is to not distrust when one or more received entries has been falsified by the other player.[14] As stated in an earlier subsection (explaining cooperative versus tournament treatments), in cooperative treatments, bonus payments are paid to both players if either one achieves accuracy. In tournament treatments, bonus payments are paid only to the first player to achieve accuracy. The combination of these two simple decisions (#Shared and Falsification) and the binary outcome coded as the variable Accuracy generate the dynamic data (i.e. this round's outcomes conditional on the previous round's outcomes) used in subsequent estimation of reaction functions. These in turn provide new empirical information about the components of the information-sharing decision process that appear to be effected by (or insensitive to) incentives.

*Feedback*

After all decisions are made in a particular round, the round concludes by announcing to each player: both players' individual accuracy, who received bonuses, and a summary of one's own information-sharing costs and total net payoff for that round (if it is selected at random to be the one for which participants are actually paid).[15] The random-round payoff technique is typically used to induce participants to play each round as if it were a one-off (thereby avoiding accumulating income and risk-diversification motives that experimental economists worry can contaminate behavioral data when participants are paid on the basis of total earnings accumulated over all experimental rounds). Nevertheless, the data indicate there is a very strong tendency to base current decisions on what others have done in the recent past.

---

[14] In Treatments A and B where information endowments consist of two entries, the choices of truthful sharing can be denoted **0**, **1**, **2** and choices when one or two shared entries are false as **F1**, **F2**, respectively. When choosing to share no entries, there is no distinction between sharing 0 true versus 0 false entries. In normal form payoff matrix in Appendices A and B, the action **0** is grouped with the truthful sharing strategies. Conditional on the other player sharing one or two entries, either true or false, (i.e., **1**, **2**, **F1**, or **F2**)**,** the player can be observed to distrust or not distrust. When the other player shares 0 database entries, the decision to distrust defaults to *not distrust.* This leads to an action space that has five elements {**0**, **1**, **2**, **F1**, or **F2**}×{*not distrust* } conditional on the other person sharing zero, and ten elements {**0**, **1**, **2**, **F1**, or **F2**}×{*not distrust,  distrust* } conditional on the other person sharing one or two.

## Descriptive Statistics and Unconditional Differences by Treatment

Table 2 presents means by treatment for #Shared (measuring the number of database entries shared on a scale of 0-2 in Treatments A and B and 0-4 in Treatments C and D) in addition to rates of Falsification and Accuracy (on 0-100 percentage-point scales). The data consist of 1,600 individual-round observations per treatment (i.e., 100 individuals observed over 16 rounds). In contrast to individual-round observations in Table 2, the primary unit of observation in the regression results presented in the next section is a *pair* of players in a single round, referred to as pair-rounds.

Comparing cooperative versus tournament incentives in Table 2 reveals sharp reductions in the mean quantity of information shared when cooperative incentives are in place: 1.616 database entries shared in cooperative Treatment A falling to 1.284 entries in tournament Treatment B (both with low-information, two-entry endowments), and 1.721 entries in C falling to 1.140 entries in D (both with high-information, four-entry endowments). The unconditional cooperative-versus-tournament differences in Table 2 easily achieve high degrees of statistical significance even after accounting for within-pair correlations among multiple observations collected from the same pair. Although mean rates of falsification are modest overall, Table 2 shows Falsification to be far higher in tournament treatments. Accuracy, interpreted as a proxy for organizational performance in tasks with demanding informational requirements, decline sharply under tournament incentives.

---

[15] Appendix 6 shows a screenshot of the individual-specific feedback that each player receives at the conclusion of each round.

Comparing low- versus high-information treatments with the same incentive scheme (i.e., A versus C and B versus D) reveals quantity-of-information effects that some may regard as counter-intuitive. Following a doubling of the number of database entries in each participant's initial endowment each round under cooperative incentives, the change in mean #Shared (comparing Treatments A and C) is statistically indistinguishable from zero, rising by less than one tenth of one database entry (from 1.616 to 1.721, or 6% with p-value = 0.277). Under tournament incentives, doubling information endowments leads to significantly less sharing, as mean #Shared declines from 1.284 to 1.140 (or 11%, p-value = 0.016).

Beyond these information effects (i.e., no increase in sharing under cooperative incentives and a decline in sharing under tournament incentives after the information endowments double), the mean accuracy rates in Table 2 reveal another curious less-is-more effect. Accuracy rates are significantly lower when players are endowed with more information. This pattern points to the possibility that larger quantities of information may paradoxically lead to declines in accuracy or organizational performance regardless of whether incentives are cooperative or individualistic.

**Results**

*Empirical Reaction Functions*

Having described the unconditional information-sharing outcomes, this section presents empirical estimates of the mean individual's reaction function under different incentive schemes conditional on previous-round information-sharing outcomes. The goal is to look for evidence about the decision processes underlying the unconditional incentive effects reported in Table 2. The reaction functions are of interest because they point to the conditionality of information-sharing strategies on cues that modulate information sharing, falsification and accuracy, and provide new descriptive information about the decision processes that generate dynamics of observed information-sharing behavior. The empirical reaction functions indicate whether incentive schemes primarily influence behavior by shifting intrinsic components of the information-sharing decision process (i.e., the components of the individual reaction function that do not depend on what other people in the organization do) or by shifting conditional responses to other people's actions (and outcomes such as Accuracy and the bonus payments that depend on Accuracy under cooperative incentives that others partially influence).

Using pair-rounds as the unit of observation for estimating empirical reaction functions that describe information-sharing dynamics in the data, the sample size is 750 based on 50 pairs per treatment observed over 15 rounds (since lagged outcomes are available only in rounds 2 through 16). Paired decisions from the previous round summarize information-sharing outcomes from the recent past and serve as the right-hand-side conditioning information in the regression models presented in this section.

*Testing Restrictions on the Seemingly Unrelated Regression Model*

The methodological issue of how to test whether regression coefficients on a particular lagged right-hand-side outcome are equal across different treatments merits some methodological discussion. The approach taken is to estimate a seemingly unrelated regression (SUR) model, which is equivalent to running separate regressions on each treatment or, equivalently, a fully interacted single-equation model pooling pair-round observations from all treatments while allowing coefficients to vary by treatment. For example, the null hypothesis that the coefficient on last period's own #Shared is equal across Treatments A and B (or that all coefficients in A and B are the same) is nested in the SUR model as an easy-to-test linear restriction. The regression results report t statistics based on robust standard errors that cluster at the level of pairs (i.e., group ids), which allows error terms in the regression model to be correlated within each paired group over the 15 pair-rounds in each treatment.

Falsification and accuracy are binary dependent variables and #Shared is a discrete integer-valued dependent variable. We ran probit/logit and ordered probit/logit models with marginal effects computed at the mean and found results qualitatively similar to the linear models reported in this section, which simplify interpretation without marginal effects depending on particular values of the right-hand side variables.

*Results for #Shared*

Table 3 presents the empirical reaction function for #Shared as a linear regression model. The coefficient 0.523 on Own_#Shared_Lag in Treatment A is interpreted as a persistence-of-sharing parameter that measures the component of the mean respondent's inter-temporal sharing rule that is independent from actions by others. Recall that the mean participant shared 1.6 database entries in Treatment A. The contribution to expected current-round sharing from previous-round sharing at the mean is therefore 1.6*0.523 = 0.84 database entries. Therefore, about half of the observed quantity of information shared in cooperative Treatment A can be explained by sheer persistence or preference for sharing (apart from the influence of the other player's decisions).

Notice, however, that in Treatment A, when a participant has made a previous decision to pass false information (i.e., the binary event of own falsification), this wipes away roughly half the contribution of intrinsic persistence in expected sharing: 1.6*0.523 – 0.426 = 0.41 (p = 0.0577, not reported in Table 3). For a person who shared one false (and zero true) database entries in the previous round, the first two coefficients in Table 3 imply a persistence factor that is indistinguishable from zero: 0.523 – 0.426 (p = 0.5340, not reported in Table 3).

Comparing coefficients on Own_#Shared_Lag across treatments, one observes remarkable consistency. The null of equal coefficients (0.523 and 0.507) in Treatments A and B cannot be rejected (p = 0.8010). Although the null of equal coefficients in Treatments C and D is rejected (p = 0.0008). But the magnitudes of persistence as measured by the coefficient on Own_#Shared_Lag in Treatments C and D do not seem to be substantially different.

Next consider responses to the other player's sharing and falsification decisions. The coefficients on Other's_#Shared_Lag and Other's_Falsified_Lag reveal interesting differences between treatments. The large-magnitude negative coefficient -0.434 in Treatment A (Table 3) implies that if the other person falsifies, then the positive effect of the other person having shared the mean number of database entries in the previous round is completely erased: 0.258*1.6 – 0.434 = -.0212 (p = 0.9283). This coefficient of -0.434 can be interpreted as evidence of a conditional punishment mechanism in the decision processes of participants: whenever the other player falsifies, the mean participant retaliates by withholding information that he or she would have otherwise shared. Based on the cross-treatment p-value of 0.0115 for Other's_Falsified_Lag (corresponding to the null of equal coefficients in Treatments A and B), retaliatory withholding of information as a form of conditional punishment appears to support higher mean rates of sharing under cooperative incentives to a substantially greater degree than under competitive incentives: -0.434 versus -0.016 in Treatment A than in Treatment B, the difference of which is statistically and economically significant. Cooperative incentives appear to trigger negative reciprocity (manifest as conditional punishment) as a mechanism for maintaining high levels of information sharing that is not observed when participants face tournament incentives. With greater endowments of information, the conditional punishment coefficients are indistinguishable from zero in Treatments C and D, reflecting statistical imprecision or sampling error associated with the fact that falsification is a relatively rare event.

One of the most noticeable results from Table 3 is the similar-sized coefficients on Other's_#Shared_Lag across all four treatments: 0.258 in A; 0.231 in B; 0.382 in C; and 0.206 in D.[16] These estimated coefficients can be interpreted as evidence of a persistent psychological mechanism employing conditional cooperation, which is little effected by different incentive schemes. This coefficient measuring conditional cooperation is significantly larger in Treatment C than in A (0.382 versus 0.258, p = 0.0011) and larger in C versus D (0.382 versus 0.206, p =

---

[16] There is a substantial literature on conditional cooperation. For example, Fischbacher, Gaechter, and Fehr (2001) find that half of the people in their experiment were willing to increase contributions to a public good in response to increases in the average contribution by others. At the same time, 30% of participants were determined free riders (contributing zero). Keser and Van Winden (2000) have found similar conditional cooperation in distribution of public good and free riders when the group of subjects played game repeatedly. In our data, the rate at which both players simultaneously shared zero was 16% based on 50 (pairs) *64 (rounds) = 3200 observations of pairs. By treatment, rates at which both people in the paired observations shared zero were: 3.4% in Treatment A, 9.4% in B, 16% in C, and 35% in D. Across all treatments and rounds, there were always more pairs with one or more individuals sharing than there were pairs with two individuals both sharing zero.

0.0442 ). Despite the larger effect in Treatment C, however, the interesting observation from our point of view is the large-magnitude effects indicating conditional cooperation based on the other player's previous-round sharing (Other's_#Shared_Lag). Across all treatments, the effect of this lagged variable on current-round own sharing (Own_#Shared) is positive and significantly different from zero by a wide margin using conservative standard errors.

If cooperative financial incentives were required to induce real people to incur the costs of passing information to an anonymous other person (as opposed to own-payoff maximizers in the theoretical model who are predicted to free-ride and share zero database entries in tournament Treatments B and D), then we would expect one of two things. If the own-payoff-maximizing model upon which the predictions of game theory are based were correct, then there would be no statistically significant reaction function conditioning on the other player's actions and outcomes that the other person influenced. Own-payoff maximizers would choose a Nash strategy and simply stick with it across all 16 rounds as long as the same incentive scheme were in place. Thus, if participants in the experimental lab were guided by game theoretic reasoning plus a random error term, the predicted result would be zero coefficients on the other player's sharing and falsification. (Without random noise, the regression would be un-estimable because of zero variation).

The other possibility, drawing on standard public goods games and the problem of free riding, would be an inexorable spiral toward zero sharing whenever the extrinsic reward for cooperating were absent (i.e., in tournament Treatments B and D). This prediction is easily rejected by the data. The coefficients on Other's_#Shared_Lag provide evidence of robust conditional cooperation regardless of differences in explicit financial incentives. Note, however, that the robustness of this conditional cooperation mechanism across treatments does not imply that treatments had no measurable effects on other components of participants' conditional reaction functions. Indeed, the four p-values for the null of all coefficients being equal across treatments indicate statistically significant differences in every pair of treatments.

Information endowments in high-information treatments (C and D) are drawn from the complete database with replacement, which introduces uncertainty not in the number of database entries (which is always 4 in Treatments C and D) but in the number of unique database entries. Including controls for the number of unique database entries reveals that when participants possess more unique information then sharing declines sharply in Treatment C. This would seem to imply that when participants are endowed with enough information that they believe their view of the organization's data is nearly complete, then they tend to "go it alone" (i.e., they stop sharing information). Although this own-unique-information effect disappears in Treatment D, the other player's possession of greater amounts of unique information significantly reduces own sharing in both Treatments C and D. This result demonstrates yet another channel through which greater quantities of information may hinder sharing even when sharing is required to achieve high rates of accuracy despite the greater endowments of information in individuals' personal possession.

Table 3 shows evidence of conditional effects of past success (i.e., accuracy) on present sharing. In competitive treatments, succeeding at achieving an accurate view of the database in the previous round increases expected #Shared by 0.142 (in Treatment B) and 0.211 (in Treatment D), both significantly different from coefficients on the same variable in cooperative treatments ($p = 0.0326$ and $0.0022$, respectively). Thus, success begets sharing even in tournament environments with highly individualistic rewards. This result suggests that individualistic incentives do not necessarily undermine sharing as long as individuals and divisions experience success with respect to their organizational objectives with sufficient frequency.

As mentioned already, the joint hypothesis that all coefficients in Treatments A and B are equal is rejected (p = 0.0419), as is the null that all coefficients in Treatments C and D are equal (p = 0.0000). This provides additional evidence that switching from tournament to cooperative incentives (or vice versa) does indeed shift the reaction functions that determine information-sharing outcomes.

A closer look at the small and statistically insignificant differences between coefficients across treatments in Table 3, however, reveals several surprises. The first two rows of coefficients represent the intrinsic or persistent component of the mean participant's reaction function (i.e., not conditional on the other player's actions). According to Table 3, cooperative incentives have no significant effect on those components of individual reaction functions that are independent of other people's decisions. Cooperative incentives, once again, do not have large effects on conditional cooperation (i.e., the coefficient on the other person's previous-round sharing). The conditional cooperation coefficients survive robustly across all treatments (as evidenced by large, nonzero and statistically significant coefficients on the other person's sharing in the previous round) although they are not significantly different across incentive schemes.

Based on Table 3, incentives appear to operate most powerfully on information sharing by activating conditional punishment. When the environment is explicitly structured to encourage cooperation, information sharing improves mostly because of the external costs that cooperative-norm-violating behavior imposes on others activate stronger punishment which leads to greater levels of sharing. Cooperative incentives distribute both rewards from success more widely *and* the pain of forgone success when groups fail to perform. Thus, cooperative incentives shift attention in ways that make others' bad behavior salient to more observers within the group or organization. This seems consistent with the observation (from post-experiment survey questionnaires) that, when paired with a free rider or falsifier, the player on the receiving end of this bad behavior (i.e., the problems of being in a group where others are withholding or falsifying information) tends to experience more anger and frustration under cooperative incentives than under tournament incentives.

The results in Table 3 suggest what appears to be an under-appreciated mechanism by which cooperative incentives succeed at shifting behavior in ways that better serve the goals of the organization. The mechanism is as described in the previous paragraph: cooperative incentives distribute losses more widely and cue punishment (likely based on a more widely distributed understanding of the group's opportunity cost when selfish or norm-violating behavior hurts the group's performance). Mutually shared awareness of the opportunity costs of poor organizational performance which are more widely distributed under cooperative incentives and the conditional punishment that individuals can expect if caught behaving non-cooperatively in an environment incentivized to reward cooperation seems to be the most important difference between the reaction functions in Table 3 under cooperative versus tournament incentives.

Participants conditionally cooperate (rather admirably) regardless of which incentive scheme they face. Only when things go wrong in cooperative reward schemes do the individual reactions to non-cooperative behavior impose a sufficiently large threat to bring about greater levels of information sharing. The models in Table 3 suggest that this mechanism is the primary reason why cooperative incentives improve quantities of information shared.

*Results for Falsification*

Falsification is a relatively rare event. Consequently, the statistical models of current-round falsification conditional on previous-round information-sharing outcomes in Table 4 have weaker explanatory power. Table 4 shows that people respond differently to other players who withhold information than they do to other players who falsify. In Treatment A, for example, if the other player falsified in the previous round, then the chance of own falsification declines. This pattern suggests that withholding information by the other player rather than falsification is the primary cue triggering retaliatory falsification in Treatment A. In Treatment B on the other hand, Table 4 indicates strong evidence that falsification by the other player in the previous round cues current-round retaliatory falsification.[17] These effects are muted in high-information Treatments C and D where there is greater uncertainty about whether falsification can be easily detected because of the positive probability that the two players have overlapping endowments (which had zero chance of occurring in Treatments A and B). This would seem to explain why mean rates of falsification are lower in Treatments C and D.

From the observed asymmetry of reactions to withholding information in cooperative treatments, and to falsification in competitive treatments, we note, once again, cooperative incentives do their work (i.e., trigger substantial shifts in behavior) by shifting what it is that individuals choose to punish. The reason, once again, would seem to be that cooperative incentives widely distribute the shared costs of forgone organizational success. Greater appreciation by many people in an organization of the opportunity cost of failing to cooperate appears to be the primary reason why cooperative incentives effectively increase cooperation at least in the context of the experiment reported in this paper.

---

[17] Fehr and Schmidt (2006, p5) seek to define the terms "cooperation," "retaliation," and "reciprocity" more narrowly than is used in this paper: "It is important to emphasize that it is not the expectation of future material benefits that drives reciprocity. Reciprocal behavior as defined above differs fundamentally from 'cooperative' or 'retaliatory' behavior in repeated interactions that is motivated by future material benefits. Therefore, reciprocal behavior in one-shot interactions is often called 'strong reciprocity' in contrast to 'weak reciprocity' that is motivated by long-term self-interest in repeated interactions."

Falsification reduces group success (i.e., Accuracy) in all treatments. Compared to tournament incentives where retaliatory falsification tends to persist more strongly, cooperative incentives appear to raise awareness about the shared cost of falsifying information. The falsification reaction functions in Table 4 lend support for earlier interpretations of the data as revealing robust conditional cooperation across all treatments contrasting with strong treatment effects cueing distinct mechanisms of punishment that support differential levels of group performance, which we turn to next.

*Results for Accuracy*

Table 5 presents expected accuracy conditional on outcomes in the previous round. Across all treatments except for C, own sharing has a positive conditional effect on accuracy and, by extension, organizational success. Similarly, the coefficients on Other's_#Shared_Lag qualitatively confirm that receiving *shared* information improves one's chances of accuracy. Own falsification is generally bad for success as is having partners that falsify.

One surprising finding in Table 5 is the large positive coefficients on Own_Accuracy_Lag. The positive coefficients of 0.197 and 0.146 (in Treatments A and B, respectively) indicate that success begets success and that self-reinforcing cycles of success in the information gathering task are present in the data. The persistence or serial correlation of success provides additional evidence that participants use adaptive information-sharing rules conditional on recent successes and failures. The finding that accuracy in the current round depends conditionally on accuracy in the previous round once again contradicts the prediction of zero correlation (round-over-round) from the theory that people play Nash strategies with an additive random noise term. According to the predictions of game-theoretic rational choice theory with an additive noise term, there should be a fixed probability of success picked up by the constant term in the regression with zero serial correlation of Accuracy if participants were playing noisy Nash strategies.

As an example of how serial correlation in Accuracy might arise, consider a pair of players whose accuracy dips during a spell in which they experiment with falsification and free riding. Following this spell of low accuracy with self-reinforcing reciprocal punishment, the participants would, thanks to random experimentation, experience a random instance of cooperative success (i.e., achieving accuracy in the information gathering task) with one or more group members having shared in the previous round. At this point, the conditional cooperation components of participants' reaction function would cue more sharing on the basis of last round's sharing (i.e., self-reinforcing conditional cooperation), leading to a spell of rounds characterized by higher rates of accuracy. Although cooperative information sharing under both incentive schemes can (and does) derail resulting in multiple rounds of retaliatory conditional punishment when a participant (either randomly or intentionally) plays uncooperatively in a cooperative treatment, it only takes one brief success at sharing and achieving accuracy for conditional cooperation amplifies past cooperation and guides this behavior to continue in future rounds with high probability.

To summarize findings from the unconditional mean differences in information sharing reported in Table 2 and the reaction functions in Tables 3-5, we note that unconditional means across treatments in Table 2 showed large increases in sharing when incentives were cooperative even though Tables 3-5 demonstrated surprising robustness of conditional cooperation under both cooperative and tournament incentives. Reinforcing this interpretation, notice in Table 5 that in both tournament Treatments B and D (in contrast to cooperative Treatments A and C) that the other person's quantity of information shared and accuracy in the previous period exert stronger influence on own accuracy in tournament treatments than in cooperative treatments. This suggests that success in groups with highly individualistic incentives depends equally if not more critically on conditional cooperation. Tables 3-5 show evidence of rather agile adaptation toward success using conditional cooperation under all incentive schemes. The explanation for large increases in sharing and reductions in falsification in cooperative treatments reported in Table 2 is the important treatment effects on negative reciprocity. Taken together, Table 2 in combination with Tables 3-5 reveal that negative reciprocity provides an additional boost of support for sharing and success. But this threat of more intense retaliation that triggers more cooperative behavior is much more pronounced under cooperative incentives.

*Order-of-Treatment Effects*

There is an important question of whether the order in which people experience different incentive schemes leads to lasting effects (i.e., historical dependence that causes measurably distinct behavioral reactions in a given incentive scheme based on exposure to different incentives that were present in the past). It may be rather ambitious to expect distinct "cultures" of cooperation and competition to emerge in a two-hour laboratory experiment, although surprisingly malleable shifts in group identity in laboratory experiments have been reported (Barkow, Cosmides and Tooby, 1992). Roughly half the experimental participants in the present study faced alternating cooperative-then-tournament treatments (A-B-C-D) while the other half faced tournament-then-cooperative sequences (B-A-D-C). Table 6 reports mean #Shared by order-of-treatment and tests the null hypothesis of no order-of-treatment effects.

We interpret these results cautiously while noting that those exposed to cooperative incentives *first* tend to share greater quantities of information in cooperative treatments. In tournament Treatment D, however, they share significantly less. This is consistent with the hypothesis that cooperative reward structures activate deep-rooted mental processes of conditional cooperation and punishment. This can lead to higher levels of cooperative information sharing when teams are functioning well and would-be non-cooperators face more intense retaliatory withholding and falsification of information whenever violations of cooperative norms occur. Investigating whether tit-for-tat conditional cooperation serves organizational objectives well by maintaining necessary levels of information sharing or perhaps generates damaging adversarial framings among divisions within an organization would of course require more context specificity describing a particular organization's goals as well as the costs and benefits of adversarial decision making. We simply note that, in some settings, negative reciprocity as a descriptive feature of information sharing behavior could, at least in theory, generate beneficial returns in the form of future cooperation. (For more on costs of negative reciprocity and the frequency of punishment, see the experimental studies of sabotage in Harbring and Irlenbusch, 2008, 2011).

The order-of-treatment effects in Table 6 are suggestive of historical dependence (Harbring, 2010). An implication of this is that organizations such as the Department of Homeland Security that have attempted to shift from previously parochial or tournament-style organizational culture toward cooperative information sharing may not realize the desired behavioral responses to the new incentives they are attempting to introduce. Or perhaps the targeted behavioral change may not occur as rapidly, or, to as great an extent, as would be the case for a new organization without its historical path that, in light of the evidence in Table 6, would appear to condition the way individuals respond to shifts in the organization's incentive scheme.

## Policy Implications?

The success of groups that learn how damaging falsification is and respond adaptively to the salience of shared costs from bad behavior when cooperative incentives are present makes use of information acquired from both failures and successes in the recent past. This is evident in the coefficients conditioning present outcomes on the previous round's sharing, falsification and accuracy. At the same time, the coefficients on own sharing are far less sensitive to incentive treatments than the own-financial-payoff-maximizing model predicts. As reported earlier, for example, sharing persists at levels substantially greater than the zero cooperation predicted by theory under tournament incentives. The data therefore provide a more subtle picture of incentive effects, suggesting multiple motives for information sharing. Consequently, while incentives are important, they are hardly the only factors in the organizational environment that influences cooperation. This raises the question of what real-world managers should take away from the empirical description of dynamic information-sharing behavior presented in the previous section.[18]

There is strong evidence of negative reciprocity as the primary mechanism that cooperative incentives activate: sharing less when others have shared less, and falsifying when others have falsified.[19] Cooperative incentives appear to amplify those forms of conditional cooperation in the context of information sharing. This could be the most important take-away (and least appreciated empirical regularity among organizational leaders, such as those who sought creation of the Department of Homeland Security with the goal of improved information sharing). Rather than people sharing more because they are being explicitly rewarded for sharing more, cooperative incentive structures in this study appear to function primarily by focusing attention on the shared losses from bad behavior more salient and activating negatively reciprocal behavior.

---

[18] One must of course apply caution when extrapolating to real-world interventions designed to facilitate information sharing. Research on the positive effects of community-of-practice technologies and face-to-face contact would, however, seem to imply potential for the large incentive effects in the lab experiments described here to complement those used in practice (Malone, Grant, Turbak, Brobst and Cohen, 1987; Lee and Whang, 2004; Mengis and Eppler, 2008; Griffith and Sawyer, 2010; Berg, Hoffrage and Abramczuk, 2010).

[19] Arguing in favor of conditional cooperation over theories of intrinsic altruism as a valid description of widespread human behavior, Frey and Meier (2004, p1717) write: "if people behave according to pure altruism theories. . .they reduce their own contribution when informed that others are already contributing."

Incentivizing divisions within an organization to share more information by distributing rewards from successes does significantly increase the quantity and quality of information shared. Under cooperative incentives, participants achieved nearly optimal levels of sharing. Under tournament incentives, however, sharing persisted to a much higher degree than the "zero sharing" prediction of theory.

Another important take-away for real-world organizations is the less-is-more effect with respect to quantity of information. Giving divisions more information leads them to share less, not more. In light of the dramatic increases in intelligence data being collected by US government agencies, the empirical results in this study raise concerns that the increased quantities of information being collected could paradoxically reduce inter-agency communication. Productive information sharing might very well be better supported by limiting the information collected by these agencies and instead working to transform organizational cultures along other dimensions.[20] The evidence suggestive of path dependence reported in the

---

[20] Woolley (2009) analyzes knowledge teams within organizations by distinguishing outcome- versus process-focused priorities. This work suggests a worthwhile direction for expanding this paper's exclusive focus on monetary incentives. Instead, extensions of the experimental design might include process-oriented or procedural shifts as incentives that might induce more behavioral response than incentive schemes based solely on monetary outcomes. Another dimension of organizational decision making that would be interesting to pursue follows Man and Lam's (2003) study of individualism/collectivism and would include treatments that vary both task complexity and autonomy. Van de Ven, Rogers, Bechara and Sun (2008) present evidence from a longitudinal study of communication and organizational change among 37 medical clinics, emphasizing the importance and challenges facing "integrative methods of open communications." While the communication analyzed in their paper clearly goes beyond data sharing as conceived in this paper's simple design, it nonetheless supports information sharing as

previous section hints at the possibility that something more fundamental than a merger of divisions could be required to succeed at wiping the slate clean and enabling improved intra-agency communication without its historical path, laden with habits of "protecting one's turf," continuing to cast a shadow over information-sharing decisions.[21]

**Conclusion**

Given the prominence of the theory that excessively individualistic incentives lead to poor information sharing which, in turn, cause organizations to fail, this paper analyzed new experimental evidence concerning how information-sharing behavior responds to cooperative versus tournament incentive schemes. The experimental design is based on an information

---

a research priority and adds breadth, in our view, to the range of interpretations of the very general (because it is simple) notion of information sharing considered in this paper.

[21] Thomas (1984) argues that disseminating statistical information to subsidiaries according to automated rules over which management has no discretionary control could help coordinate decisions among a multi-location organization. The effects of automation and elimination of discretion could be implemented as further treatments using the information-sharing game in future research. Mitsuhashi (2003) discusses how alliances that facilitate information sharing may paradoxically limit heterogeneity of the information one gathers, because alliances' information sets are correlated. The information-sharing subsidiaries in our experiment comprise only a single alliance and their information sets are perfectly (in Treatments A and B) or near-perfectly (in Treatments C and D) complementary. The possibility that alliances cluster around other dimensions of similarity where that information is least valuable for the organization is yet another direction for extending the information-sharing game in future research: allowing for multiple alliances and variation in the correlatedness of players' information endowments.

sharing game that makes clear predictions about the effects of different incentive schemes on quantities of information shared, falsification, and accuracy as a proxy for organizational performance.

The data show that cooperative incentives in the lab do succeed at increasing quantities of shared information. The mechanism for this, however, does not work by shifting intrinsic preferences for, or the persistence of, sharing. Rather, the empirical reaction functions reported in this paper show that cooperative incentives operate primarily by intensifying negative reciprocity in the form of retaliatory withholding of information and falsification.

Distributing the rewards of success more widely also distributes the pain of failure more widely. Thus, cooperative incentives exert influence on sharing behavior primarily by making the costs of others' under-sharing and falsification more salient, activating willingness to undertake conditional punishment, and thereby enforce higher levels of cooperative sharing. The pessimistic theoretical prediction of zero sharing under tournament rewards is easily rejected by the data. And although there is measurably less sharing under tournament incentives, a rather large number of teams continue to conditionally cooperate and conditionally punish to sustain impressive levels of sharing and accuracy. These robust cooperators deviate from one-shot best-response strategies in the theoretical model and, in so doing, perform better than they would have by Nash-best-responding. Although falsification is never a theoretical best response when players believe they are playing with own-monetary-payoff-maximizers, tournament incentives generate significantly higher rates of falsification. Falsification was not absent, however, under cooperative incentives. As a punishment mechanism to enforce future cooperation, it is possible that falsification may have played a productive role as punishment in support of adaptation toward cooperation in some teams.

Rates of accuracy in the information aggregation task increased 15 to 20 percentage points under cooperative incentives relative to base rates of around 50% under tournament incentives. Endowing participants with more information failed to increase (or reduced) sharing: High-information treatments also led to weak and/or counter-intuitive effects on accuracy. In tournament conditions, more information raised accuracy by less than 7 percentage points and reduced accuracy under cooperative incentives. Increased quantities of information do not appear to activate, and may detract from, effective information sharing. While the increased information content of larger datasets is understandably attractive, the less-is-more effects reported here (regarding quantities of information in one's possession and effective sharing of that information) raise questions about the information-sharing strategies of some organizations based on an apparent philosophy of more-is-better.

Butler, J. (1999). Trust expectations, information sharing, climate of trust, and negotiation effectiveness and efficiency. *Group and Organization Management 24* (2), 217-238.

Centers for Disease Control and Prevention. (2012). *CAPABILITY 6: Information Sharing,* Retrieved Jan 10, 2013, from *www.cdc.gov/phpr/capabilities/capability6.pdf*

Cabrera, A. and E. Cabrera (2002). Knowledge-sharing dilemmas. *Organization studies 23* (5), 687-710.

Campbell, D. and D. Furrer (1995). Goal setting and competition as determinants of task performance. *Journal of Organizational Behavior 16* (4), 377-389.

Cason, T. (1994). The impact of information sharing opportunities on market outcomes: an experimental study. *Southern Economic Journal 61*, 18-39.

Chalos, P. and S. Haka (1990). Transfer pricing under bilateral bargaining. *Accounting Review 65*, 624-641.

Chang, L., M. Cheng, and K. Trotman (2008). The effect of framing and negotiation partners objective on judgments about negotiated transfer prices. *Accounting, Organizations and Society 33* (7), 704-717.

Chen, F. (2003). Information sharing and supply chain coordination. In de Kok, T. and S. Graves (Eds.), *Handbooks in Operations Research and Management Science: Supply Chain Management,* Vol. 11, Elsevier, pp. 341-421.

Cheng, M. and C. Hsieh (2009). Transfer price negotiation in the presence of unequal bargaining power: The effect of a peer evaluation scheme on inter-divisional profit distribution. *Australian Accounting Review 19* (3), 195-206.

Connelly, C., D. Zweig, J. Webster, and J. Trougakos. (2012). Knowledge hiding in organizations. *Journal of Organizational Behavior 33* (1), 64-88.

Dawes, S. (1998). Interagency information sharing: Expected benefits, manageable risks. *Journal of Policy Analysis and Management 15* (3), 377-394.

Dejong, D., R. Forsythe, J. Kim, and W. Uecker. (1989). A laboratory investigation of alternative transfer pricing mechanisms. *Accounting, Organizations and Society 14* (1-2), 41-64.

Dyer, J.H. and W. Chu. (2003). The role of trustworthiness in reducing transaction costs and improving performance: Empirical evidence from the United States, Japan, and Korea. *Organization Science 14* (1), 57-68.

Eby, L. and G. Dobbins. (1997). Collectivistic orientation in teams: An individual and group-level analysis. *Journal of Organizational Behavior 18*, 275-295.

Ensley, M. and C. Pearce. (2001). Shared cognition in top management teams: Implications for new venture performance. *Journal of Organizational Behavior 22* (2), 145-160.
39

Nalbantian, H. and A. Schotter. (1997). Productivity under group incentives: An experimental study. *American Economic Review 87* (3), 314-341.

Quigley, N., P. Tesluk, E. Locke, and K. Bartol. (2007). A multilevel investigation of the motivational mechanisms underlying knowledge sharing and performance. *Organization Science 18* (1), 71-88.

Ravenscroft, S., S. Haka, and P. Chalos. (1993). Bargaining behavior in a transfer pricing experiment. *Organizational Behavior and Human Decision Processes, 55* (3), 414-443.

Rotolo, D. and A.M. Petruzzelli. (forthcoming). When does centrality matter?: Scientific productivity and the moderating role of research specialization and cross-community ties. *Journal of Organizational Behavior*.

Sahin, F. and E. Robinson. (2007). Flow coordination and information sharing in supply chains: review, implications, and directions for future research. *Decision Sciences 33* (4), 505-536.

Salas, E. and J. Cannon-Bowers. (2001). The science of training: A decade of progress. *Annual Review of Psychology 52* (1), 471-499.

Sawers, K. and W. Liao. (2005). An experimental comparison of transfer pricing methods under high and low private information, AAA Management Accounting Section 2006 Meeting Paper.

Thomas, R. (1984). Statistical information systems and management1. *Organization Studies 5* (4), 345-358.

Tsai, W. (2002). Social structure of coopetition within a multiunit organization: Coordination, competition, and intraorganizational knowledge sharing. *Organization science 13* (2), 179-190.

United States, Information Sharing Environment. (n.d.). *Common information sharing standards*. Retrieved from website: http://www.ise.gov/common-information-sharing-standards.

United States Congress, Senate. (Dec. 7, 2004). Conference Report on S. 2845, Intelligence Reform and Terrorism Prevention Act of 2004. (H. Rept. 108-796) (Page H10930-H10993). Retrieved from website: http://www.fas.org/irp/congress/2004_rpt/index.html

Van de Ven, A., R. Rogers, J. Bechara, and K. Sun. (2008). Organizational diversity, integration and performance. *Journal of Organizational Behavior 29* (3), 335-354.

Vives, X. (1984). Duopoly information equilibrium: Cournot and Bertrand. *Journal of Economic Theory 34* (1), 71-94.

Vora, D. and T. Kostova. (2007). A model of dual organizational identification in the context of the multinational enterprise. *Journal of Organizational Behavior 28* (3), 327-350.

Wageman, R. and G. Baker. (1997). Incentives and cooperation: The joint effects of task and reward interdependence on group performance. *Journal of Organizational Behavior 18*(2), 139-158.

Wagner, J.A., S.E. Humphrey, C.J. Meyer, and J.R. Hollenbeck. (2012). Individualism-collectivism and team member performance: Another look. *Journal of Organizational Behavior 33* (7), 946-963.

Whitley, R. (2003). The institutional structuring of organizational capabilities: the role of authority sharing and organizational careers. *Organization Studies 24* (5), 667-695.

Woolley, A.W. (2009). Putting first things first: Outcome and process focus in knowledge work teams. *Journal of Organizational Behavior*, *30* (3), 427-452.

Zhou, H. and W. Benton. (2007). Supply chain practice and information sharing. *Journal of Operations Management 25* (6), 1348-1365.

43